\documentclass[pra,showpacs,preprintnumbers,amsmath,aps]{revtex4}
\usepackage{graphics}

\def\be{\begin{equation}}
\def\ee{\end{equation}}
\def\bea{\begin{eqnarray}}
\def\eea{\end{eqnarray}}
\def\bse{\begin{subequations}}
\def\ese{\end{subequations}}
\def\bma{\begin{mathletters}}
\def\ema{\end{mathletters}}
\def\C{\hbox{$\mit I$\kern-.6em$\mit C$}}

\begin{document}

\title{Distinguishability, classical information of quantum operations}
\author{Dong Yang}
\email{dyang@ustc.edu.cn}
\affiliation{Department of Modern Physics, University of Science and Technology of China, Hefei, Anhui 230026, People's Republic of China}

\date{\today}

\begin{abstract}
A basic property of distinguishability is that it is non-increasing under further quantum operations. Following this, we generalize two measures of distinguishability of pure states---fidelity and von Neumann entropy, to mixed states as self-consistent measures. Then we extend these two measures to quantum operations. The information-theoretic point of the generalized Holevo quantity of an ensemble of quantum operations is constructed. Preferably it is an additive measure. The exact formula for $SU(2)$ ensemble is presented. With the aid of the formula, we show Jozsa-Schlienz paradox that states as a whole are less distinguishable while all pairwise are more distinguishable in an ensemble of quantum states, also occurs in an ensemble of quantum operations, even in the minimal dimensional case $SU(2)$ ensemble.
\end{abstract}

\pacs{03.67.-a, 03.65.Ta}

\maketitle

\section{Introduction}
Quantum nonorthogonality is one of the basic features of quantum mechanics. Unlike the distinct states in classical physics, it is impossible to discriminate perfectly between nonorthogonal states if only one copy is provided. Motivated by this simple fact, there have been attempts to define measures to quantify the degree of distinguishability between quantum states. A well-known measure is the fidelity between two quantum states \cite{Jozsa3} that describes: the more similar, the less distinguishable they are. Recently, von Neumann entropy is suggested to measure the distinguishability for an ensemble of pure states from the information-theoretic point \cite{Jozsa1}: the more classical information they can communicate, the more distinguishable they are. Indeed, von Neumann entropy is the classical information capacity communicated by an ensemble of pure states \cite{Hausladen}. Quantum states and quantum dynamics are two parts of quantum mechanics. Both quantum states and dynamics are fundamental physical resources \cite{Nielsen1}. A general quantum dynamical process is described by a quantum operation, a completely positive trace-preserving linear map (CPT) \cite{Choi,Schumacher1}. Quantum nonorthogonality occurs not only in quantum states but also in quantum operations. It is possible that two operations can not be discriminated exactly if only one operation is performed. Distinguishability between two unitary operations is discussed in \cite{Childs,Acin1,D'Ariano1}. However, little is known about the distinguishability of general quantum operations. Motivated by this, we try to quantify the distinguishability of quantum operations.

How can one introduce a measure of distinguishability for quantum operations? A natural approach exists if two key points are noticed. One is the basic property of distinguishability that it is non-increasing under further quantum operations and the other is that distinguishability of quantum operations can be defined based on distinguishability of quantum states. More specifically speaking, distinguishability of pure states is extended to that of mixed states, and then distinguishability of quantum operations is proposed as the largest distinguishability of the output states (mixed states). In this paper, we discuss two measures of distinguishability: one is fidelity and the other is Holevo quantity \cite{Kholevo}. The rest of the paper is organized as follows. Section \ref{se:map} recalls the notions of CPT map and properties of Holevo quantity that is useful to study distinguishability. Section \ref{se:states} first reviews two measures of distinguishability of pure states---fidelity and von Neumann entropy, then generalizes to those of mixed states as self-consistent measures. Section \ref{se:operations} explores the corresponding measures for distinguishability of quantum operations. Several propositions are proved and the information-theoretic explanation of Holevo quantity is constructed. Further the formula to calculating the distinguishability of $SU(2)$ ensemble is provided. Employing the formula, a counter-intuitive phenomenon is discovered analogous to Jozsa-Schlienz paradox in \cite{Jozsa1}. Section \ref{se:remarks} concludes with summary.

\section{CPT map and Holevo quantity}
\label{se:map} Before discussing distinguishability, we briefly review the CPT map and Holevo quantity that are used in the latter sections.

\subsection{CPT Map}
A general quantum states is described by a positive Hermite operator with trace one. A general quantum dynamical process is described by a quantum operation, a completely positive trace-preserving linear map (CPT)\cite{Choi,Schumacher1}, $\rho_{out}={\cal E}(\rho_{in})$. 'Linear' means ${\cal E}(p_{1}\rho_{1}+p_{2}\rho_{2})=p_{1}{\cal E}(\rho_{1})+p_{2}{\cal E}(\rho_{2})$. 'Trace-preserving' means $tr {\cal E}(\rho)=tr \rho$. 'Positive' means ${\cal E}(\rho)\ge 0$ if $\rho\ge 0$. 'Completely positive' means ${\cal E}\otimes I$ is positive where $I$ is the identity map on any dimensions. There are two useful representations of a CPT map ${\cal E}$. One is the operator-sum representation, ${\cal E}(\rho)=\sum_{i}A_{i}\rho A_{i}^{\dagger}$, where $A_{i}$ are operators satisfying $\sum_{i}A_{i}^{\dagger} A_{i}=I$. The other is unitary representation: a CPT map can be written as a unitary evolution on an extended system $AB$ followed by a partial trace over $B$, ${\cal E}^{A}(\rho^{A})=tr_{B}U^{AB}\rho^{A}\otimes(|0\rangle\langle 0|)^{B}U^{AB\dagger}$. For more details and the relation between the two representations, the reader is refereed to \cite{Choi,Schumacher1}.

\subsection{von Neumann Entropy and Holevo quantity}
Von Neumann entropy of a density operator is defined as $S(\rho)=-tr\rho\log\rho$, where logarithm is to base $2$. It has a remarkable property known as strong subadditivity inequality \cite{Wehrl}, $ S(\rho_{12})+S(\rho_{23})\ge S(\rho_{123})+S(\rho_{2})$, with any tripartite state $\rho_{123}$ on Hilbert space ${\cal H}_{1}\otimes {\cal H}_{2}\otimes {\cal H}_{3}$.

An ensemble of mixed states is denoted as $\{\rho_{i}, p_{i}\}$ where $\rho_{i}$ are quantum states and $p=\{p_{i}\}$ is a probability distribution satisfying $p_{i}\ge 0, \sum_{i}p_{i}=1$. Holevo quantity for an ensemble of mixed states is defined as, \be \chi(\{\rho_{i}, p_{i}\})=S(\sum p_{i}\rho_{i})-\sum p_{i}S(\rho_{i}). \ee For an ensemble of pure states, Holevo quantity is reduce to von Neumann entropy. Holevo quantity has two properties \cite{Lindblad,Uhlmann,Hayden}.

{\bf Proposition}: Holevo quantity is non-increasing under trace operation and under quantum operation,
\bse
\bea
\chi(\{\rho^{AB}_{i}, p_{i}\})\ge \chi(\{\rho^{A}_{i}, p_{i}\}),\label{trace}
\\
\chi(\{\rho_{i}, p_{i}\})\ge \chi(\{{\cal E}(\rho_{i}), p_{i}\}), \label{operation}
\eea
\ese
where $\rho^{A}_{i}=tr_{B}\rho^{AB}_{i}$ and ${\cal E}$ is a CPT map.

The proof comes from the strong subadditivity inequality of von Neumann entropy \cite{Lindblad,Uhlmann,Hayden}.

\section{Distinguishability of Quantum States}
\label{se:states} Now we investigate the distinguishability measure of quantum states. Intuitively, the word {\it distinguishability} means to what extent things are distinct from each other. As a meaningful measure of distinguishability among the objects, the quantity cannot be increased by the same further processing, otherwise we are in dilemma that the quantity tends to infinity that is clearly meaningless. In quantum mechanics, the further processing is quantum operation described by CPM. So we argue that a reasonable measure of distinguishability defined in quantum mechanics should satisfy the constraint: {\it a measure of distinguishability is non-increasing under any further quantum operation.} We will see how this constraint leads us to define a self-consistent measure of distinguishability.

\subsection{Fidelity}
From the intuition that the more alike the less distinguishable, distinguishability of two states is measured by fidelity. First it is defined on two pure states, then is extended to mixed ones under the constraint.

Suppose two pure nonorthogonal states $|\phi\rangle$ and $|\psi\rangle$, the overlap or fidelity is defined as $F(\phi, \psi)=|\langle \phi|\psi\rangle|$ and measures to what extent the two states behave alike. Indeed, $|\langle \phi|\psi\rangle|^{2}$ is the probability that $|\phi\rangle$ passes the test of 'being state $|\psi\rangle$'. Note that the fidelity is invariant under unitary evolution which means that the two states behave the same alike as the beginning if they evolve under the same unitary dynamics. The more alike, the less distinguishable they are. So fidelity can be used to mark distinguishability. However, before that we should check whether fidelity is non-decreasing under quantum operation as distinguishability is non-increasing under quantum operation. Under general quantum dynamics, pure states evolve into mixed states. As a natural requirement, it is necessary to define fidelity between two mixed states such that it could be reduced to the pure case when they are pure. Note that mixed states can be purified to entangled pure states (purification) with the aid of an auxiliary subsystem, and can be regarded as partially tracing the auxiliary subsystem over the purification. It is possible to define fidelity of two mixed states as that of purifications. Given a mixed state, purification is not unique---infinite purifications exist. How do we select a pair of purifications for definition? The answer also comes from the constraint that distinguishability is non-increasing under quantum operations. As partial-tracing is also a quantum operation, only one way is left to define fidelity of mixed states as,
\be
F(\rho_{1},\rho_{2})=\max|\langle\Phi_{1}|\Phi_{2}\rangle|,\label{F}
\ee
where the maximum is taken over all purifications $|\Phi_{1}\rangle,|\Phi_{2}\rangle$. The physical meaning is that $F(\rho_{1},\rho_{2})$ is the worst distinguishability between purifications. This definition is also accordance with the meaning of distinguishability: the more we learned, the better we can distinguish. But does $F(\rho_{1},\rho_{2})$ satisfy the requirement of distinguishability? Indeed, the definition of (\ref{F}) is self-consistent, which means that $F(\rho_{1},\rho_{2})$ is non-decreasing under further quantum operation. Indeed, Jozsa proved the following proposition\cite{Jozsa3}.

{\bf Proposition}: $F(\rho_{1},\rho_{2})$ is non-decreasing under tracing subsystem operation and under quantum operation, \bse \bea F(\rho_{1}^{A},\rho_{2}^{A})\ge F(\rho_{1}^{AB},\rho_{2}^{AB}),
\label{st}\\
F({\cal E}(\rho_{1}),{\cal E}(\rho_{2}))\ge F(\rho_{1},\rho_{2}), \label{so} \eea \ese where $\rho_{i}^{A}=tr_{B}\rho_{i}^{AB}$.

It is further showed \cite{Jozsa3} that$F(\rho_{1},\rho_{2})=tr[(\sqrt{\rho_{1}}\rho_{2}\sqrt{\rho_{1}})^{1/2}]$,where $\{tr[(\sqrt{\rho_{1}}\rho_{2}\sqrt{\rho_{1}})^{1/2}]\}^{2}$ is introduced as 'transition probability' for mixed states by Uhlmann in \cite{Uhlmann2}. For more details about $F(\rho_{1},\rho_{2})$, the readers are refereed to \cite{Jozsa3}.

\subsection{von Neumann Entropy and Holevo quantity}
The distinguishability measure by fidelity is defined for two states. How can we define the measure for an ensemble of states? From the information-theoretic point that the more distinguishable of a set of states are, the more information they can communicate, Jozsa and Schlienz \cite{Jozsa1} proposed that von Neumann entropy can be used to quantify distinguishability of an ensemble of pure states, where the distinguishability measure of $E=\{|\phi_{i}\rangle, p_{i}\}$ is defined as $D(E)=S(\sum p_{i}|\phi_{i}\rangle\langle\phi_{i}|)$. Indeed, von Neumann entropy can be explained as the classical information capacity communicated by the ensemble of pure states \cite{Hausladen}. As a self-consistent definition, a distinguishability measure should be defined on mixed states because the measure should be non-increasing under further quantum operation and pure states generally evolve into mixed ones. An explicit generalization of von Neumann entropy is Holevo quantity in the mixed case $E=\{\rho_{i}, p_{i}\}$,
\be
D(E)=\chi(\{\rho_{i}, p_{i}\}).
\ee

The Holevo quantity is non-increasing under further quantum operation. More importantly, it does measure the classical information capacity for an ensemble of mixed states \cite{Schumacher3}. So the Holevo quantity is indeed a self-consistent measure based on the information-theoretic point.

{\it Remarks}: $(1)$ Intuitively, the more distinguishable each pair of quantum states, the larger distinguishability of the ensemble. However, a counter-intuitive fact known as Jozsa-Schlienz paradox is found in \cite{Jozsa1}, where the states becomes all pairwise more distinguishable while the distinguishability of the ensemble measured by von Neumann entropy decreases. This phenomenon shows that distinguishability measured by von Neumann entropy is a global property of an ensemble of pure states and not an accumulative local property of pairs of constituent states \cite{Jozsa1}. $(2)$ From the information-theoretic explanation, von Neumann entropy is a quantity in the asymptotic meaning \cite{Schumacher1,Hausladen}, in which physical quantities is generally defined by the regularization form, i. e. $\lim_{n\to \infty}D(E^{\otimes n})/n$. Just because von Neumann entropy and Holevo quantity are additive over tensor product of ensembles, the regularization is reduced to one copy. We are faced with the regularization problem when we define the distinguishability measure of operations.

\subsection{Orders of Different Measures}
Both fidelity and von Neumann entropy (Holevo quantity) can be used to measure the distinguishability of an ensemble with two states. For pure states, distinguishability measured by fidelity and von Neumann entropy give the same order, i. e. if ensemble $E_{1}$ is more distinguishable than $E_{2}$ measured by fidelity, then it is also the case when measured by von Neumann entropy and vice versa. This can be easily seen that von Neumann entropy is monotonously dependent on the inner product of two states. However, for general mixed states it is possible that $E_{1}$ is more distinguishable than $E_{2}$ measured by fidelity while $E_{1}$ is less distinguishable than $E_{2}$ by Holevo quantity. A simple situation is: suppose two ensembles $E_{1}=\{\rho_{1},\rho_{2}, p_{1},p_{2}\}$ and $E_{2}=\{|\phi_{1}\rangle,|\phi_{2}\rangle, p_{1},p_{2}\}$ where $|\phi_{1}\rangle$ and $|\phi_{2}\rangle $ are the optimal purifications to achieve $F(\rho_{1},\rho_{2})=|\langle\phi_{1}|\phi_{2}\rangle|$. From ${\it Eq.}(\ref{trace})$, $\chi(E_{1})\le \chi(E_{2})$ holds. For generic mixed states, $\chi(E_{1})$ is strictly less than $\chi(E_{2})$. As von Neumann entropy is continuously dependent on $\langle\phi_{1}|\phi_{2}\rangle$, $|\langle\phi_{1}|\phi_{2}\rangle|$ can be made a little larger by deforming $|\phi_{1}\rangle$ and $|\phi_{2}\rangle$ while the inequality $\chi(E_{1}) < \chi(E_{2})$ still holds. So distinguishability measured by fidelity is not always inconsistent with that by Holevo quantity. The fact that different measures give different orders of distinguishability implies that a particular measure just reflects a particular property of distingushability in quantum states.

\section{Distinguishability of Quantum Operations}
\label{se:operations} {\bf Definition}: An ensemble of quantum operations is defined as an ensemble $\{{\cal E}_{i}, p_{i}\}$, where ${\cal E}_{i}$ are CPT maps and $p=\{p_{i}\}$ is a probability distribution. When all ${\cal E}_{i}$ are unitary operators $U_{i}$, we call the ensemble $\{U_{i},p_{i}\}$ unitary ensemble.

Suppose we are given one of black boxes that perform operations ${\cal E}_{i}$ with probability $p_{i}$ and required to identify which operation the given box performs. Of course, if the black box can be inquired infinite times, that is to say the same operation can be performed infinitely, we can identify what the operation is by quantum operation tomography \cite{D'Ariano3}. If we can inquire the box only once, to what extent can we tell which operation the box performs? But before that, what is the meaning of 'extent'? Therefore, it is necessary to define the measure of distinguishability--- the degree of distinguishability between quantum operations. To distinguishing quantum operations, the only way is to input a state to the black box and the evolution is inferred by the output state. So distinguishability for an ensemble of quantum operations is reduced to distinguishability of the ensemble of output states. As there exist different ensembles of output states for different input states, distinguishability of the ensemble of quantum operations is defined as the maximal distinguishability of the ensemble of output states. The corresponding input state is the optimal one to distinguish the quantum operations as well as possible. Note also that given a CPT ${\cal E}_{i}$, we are allowed to perform any operation of the form ${\cal E}_{i}\otimes I $ where $I$ is the identity operator acting on any dimensions. So it is possible that an entangled state including an auxiliary subsystem is better for distinguishing quantum operations. An explicit example is that $\{I,\sigma_{x},\sigma_{y},\sigma_{z}\}$ can be distinguished exactly by a maximally entangled state $|\phi\rangle=1/\sqrt{2}(|00\rangle+|11\rangle)$, where $\sigma_{x},\sigma_{y},\sigma_{z}$ are Pauli operators. Here $\{{\cal E}_{i}, p_{i}\}$ is just a brief notation that actually means $\{{\cal E}_{i}\otimes I, p_{i}\}$.

In this section, parallel to distinguishability of quantum states measured by fidelity and Holevo quantity, the corresponding distinguishability of quantum operations is studies respectively.

\subsection{Fidelity}
In this subsection, we discuss fidelity of two quantum operations. In \cite{Acin1}, distinguishability between two unitary operators $U_{1},U_{2}$ is defined by the minimal fidelity, $ F(U_{1},U_{2})=\min_{|\phi\rangle}F(U_{1}\otimes I|\phi\rangle,U_{2}\otimes I|\phi\rangle)$. From the same reason as quantum states, a self-consistent measure should be defined on general quantum operations.

{\bf Definition} The fidelity of two quantum operations between ${\cal E}_{1}$ and ${\cal E}_{2}$ is defined as
\be
F({\cal E}_{1},{\cal E}_{2})=\min_{\rho}F({\cal E}_{1}\otimes I(\rho),{\cal E}_{2}\otimes I(\rho)). \label{DF}
\ee
Before we prove that $F({\cal E}_{1},{\cal E}_{2})$ is non-decreasing under further quantum operation, we show that minimization can be obtained over a pure state entangled with a finite-dimensional system.

{\bf Lemma 1} Given two quantum operations ${\cal E}_{1}$ and ${\cal E}_{2}$ acting on $d$-dimensional system $A$. $F({\cal E}_{1},{\cal E}_{2})$ can be achieved by minimization over pure states entangled with a $d$-dimensional auxiliary system $B$ at most. That is \be F({\cal E}_{1},{\cal E}_{2})=\min_{|\phi\rangle\in d\otimes d}F({\cal E}_{1}\otimes I_{d}(\phi),{\cal E}_{2}\otimes I_{d}(\phi)) \ee {\it Proof.} For a general mixed state $\rho^{AB}$ on $d\otimes n$-dimensional space, it can be always purified into a pure state $|\Phi\rangle^{ABC}$ with another auxiliary system $C$, where $|\Phi\rangle^{ABC}$ can be written in its Schmidt decomposition splitting between $A$ and $BC$, \be |\Phi\rangle^{ABC}=\sum_{i=0}^{d-1}\sqrt{\lambda_{i}}|i\rangle^{A}\otimes |e_{i}\rangle^{BC}, \ee in which $|e_{i}\rangle^{BC}$ are orthogonal states spanning $d$-dimensional space. From $Eq.$(\ref{st}), we can get \bea F({\cal E}_{1}^{A}\otimes I^{B}(\rho^{AB}),{\cal E}_{2}^{A}\otimes I^{B}(\rho^{AB}))&=&F(tr_{C}{\cal E}_{1}^{A}\otimes I^{BC}(\Phi^{ABC}),tr_{C}{\cal E}_{2}^{A}\otimes I^{BC}(\Phi^{ABC}))
\nonumber\\
&\ge& F({\cal E}_{1}^{A}\otimes I^{BC}(\Phi^{ABC}),{\cal E}_{2}^{A}\otimes I^{BC}(\Phi^{ABC})). \eea Notice that $\Phi^{ABC}$ can be transform to the standard form $\sum_{i=0}^{d-1}\sqrt{\lambda_{i}}|i\rangle^{A}\otimes |i\rangle^{BC}$ under unitary operator $I^{A}\otimes U^{BC}$ without varying $F({\cal E}_{1}^{A}\otimes I^{BC}(\Phi^{ABC}),{\cal E}_{2}^{A}\otimes I^{BC}(\Phi^{ABC}))$. So the minimization can be considered over pure states in the $d\otimes d$-dimensional Hilbert space.

{\bf Proposition 1} $F({\cal E}_{1},{\cal E}_{2})$ is non-decreasing under further quantum operation, \be F({\cal N\circ E}_{1},{\cal N\circ E}_{2})\ge F({\cal E}_{1},{\cal E}_{2}), \ee where ${\cal N}$ is a general quantum operation.

{\it Proof.} The proof is immediately obtained from $Eq.$(\ref{so}), \be F({\cal N\circ E}_{1}\otimes I_{d}(\phi),{\cal N\circ E}_{2}\otimes I_{d}(\phi))=F({\cal N}\otimes I({\cal E}_{1}\otimes I_{d}(\phi)),{\cal N}\otimes I({\cal E}_{2}\otimes I_{d}(\phi))\ge F({\cal E}_{1}\otimes I_{d}(\phi),{\cal E}_{2}\otimes I_{d}(\phi)). \ee

The problem of distinguishing quantum operations is different from that of quantum states though they seem much alike. In \cite{Acin1}, a distinct property is demonstrated that for any two unitary operations $U_{1}$ and $U_{2}$ there always exists a finite number $N$ such that $U_{1}^{\otimes N}$ and $U_{2}^{\otimes N}$ are perfectly distinguishable although they were not in the single-copy case. Recall that it is not the case for two nonorthogonal states. If two states are not discriminated perfectly with one copy, they are not with any finite copies. The input state entangled with different subsystems on which every $U$ performs improves the distinguishability between $U_{1}^{\otimes N}$ and $U_{2}^{\otimes N}$. The result also implies that any number unitary operations can be  exactly distinguished if enough finite copies are provided. An upper bound on copies is $N=\sum_{k=1}^{m-1}N^{(k)}$, where $N^{(k)}$ is the $k$-th number in the decreasing sequence of $\{N_{ij},i\ne j\}$ and $N_{ij}$ is the minimal copies that are required to distinguish $U_{i}$ and $U_{j}$ from the set $\{U_{i},i=1,\cdots,m\}$. The reason is that we can always use $N^{(k)}$ copies to exclude one possible and if the possible is excluded, the number of required copies does not appear in later exclusion. So we can exclude all possibilities arbitrarily. For example, let us denote the possible operation $U_{i}$. First, we use $N_{12}$ copies to exclude the possibility of $U_{1}$ or $U_{2}$ by proper input state. If the outcome is in favor of $U_{1}$, the possibility of $U_{2}$ is excluded. Then we are left with $m-1$ possibilities and continue in this way until there is left only one possible. Here we don't optimize the problem. Indeed, much better upper bound can be obtained. Is it true that general operations can be perfectly distinguished with finite copies? The answer is negative. In the following, we give two instances.

A CPT map is called entanglement breaking if ${\cal E}^{A}\otimes I^{B}(\rho^{AB})$ is always separable for any $\rho^{AB}$. A CPT map is entanglement-breaking if and only if it can be written in the form \cite{HSR}
\be
{\cal E}(\rho)=\sum_{i}|\phi_{i}\rangle\langle\phi_{i}|tr (|\psi_{i}\rangle\langle\psi_{i}|\rho),
\ee
where $|\phi_{i}\rangle$ are normalized and $|\psi_{i}\rangle$ are not necessarily normalized but satisfy $\sum_{i}|\psi_{i}\rangle\langle\psi_{i}|=I$.

{\bf Ex 1}: If two unitary operations $U_{1},U_{2}\in SU(d)$ are not perfectly distinguishable in the single-copy case, then two EB ${\cal E}_{1}=U_{1}\circ{\cal E},{\cal E}_{2}=U_{2}\circ{\cal E}$ are not perfectly distinguishable in any finite-copy case, in which ${\cal E}$ is an EB map.

{\it Proof.}
\be
{\cal E}_{1(2)} (\rho_{1(2)})=\sum_{i}U_{1(2)}|\phi_{i}\rangle\langle\phi_{i}|U_{1(2)}^{\dagger}\otimes|tr (|\psi_{i}\rangle\langle\psi_{i}|\rho_{1(2)}),
\ee

Here $\rho_{1(2)}$ are bipartite states with ancilla system. Since $U_{1}|\phi_{i}\rangle$ is not orthogonal to $U_{2}|\phi_{i}\rangle$, if ${\cal E}_{1} (\rho_{1})\perp {\cal E}_{2} (\rho_{2})$, then $tr (|\psi_{i}\rangle\langle\psi_{i}|\rho_{1})\perp tr (|\psi_{i}\rangle\langle\psi_{i}|\rho_{2})$. As $\{|\psi_{i}\rangle\langle\psi_{i}|\}$ is a generalized measurement, it means that $\rho_{1}\perp\rho_{2}$.

\bea
\rho_{1(2)}&=&{\cal E}_{1(2)}\otimes{\cal E}_{1(2)}\circ{\cal E}_{1(2)}\cdots{\cal E}_{1(2)}(\phi)\nonumber\\
&=&({\cal E}_{1(2)}\otimes I\otimes\cdots\otimes I)
\circ (I\otimes{\cal E}_{1(2)}\otimes I\cdots\otimes I)
\circ  (I\otimes{\cal E}_{1(2)}\otimes I\cdots\otimes I)
\circ \cdots(I\otimes I\otimes\cdots\otimes {\cal E}_{1(2)})(\phi)
\eea
By induction, the two operations must be perfectly distinguishable in the single-copy case. However this means $U_{1}$ and $U_{2}$ can be distinguished with one copy that contradicts with the supposition.

{\bf Ex 2}: If two distinct EB maps ${\cal E}_{1},{\cal E}_{2}$, ${\cal E}_{1(2)}(\rho)=\sum_{i}|\phi_{i1(2)}\rangle\langle\phi_{i1(2)}|tr (|\psi_{i1(2)}\rangle\langle\psi_{i1(2)}|\rho)$ satisfy $|\langle\phi_{i1}|\phi_{j2}\rangle|\ne 0$ for all $i,j$, then they are not perfectly distinguishable in any finite-copy case.

{\it Proof.}
\be
{\cal E}_{1(2)} (\rho_{1(2)})=\sum_{i}|\phi_{i1(2)}\rangle\langle\phi_{i1(2)}|\otimes tr (|\psi_{i1(2)}\rangle\langle\psi_{i1(2)}|\rho_{1(2)}),
\ee
If ${\cal E}_{1} (\rho_{1})$ is orthogonal to ${\cal E}_{2} (\rho_{2})$, then $ tr (|\psi_{i1}\rangle\langle\psi_{i1}|\rho_{1})$ is orthogonal to $ tr (|\psi_{j2}\rangle\langle\psi_{j2}|\rho_{2})$. So $tr\rho_{1}$ is orthogonal to $tr\rho_{2}$ that means $\rho_{1}$ is orthogonal to $\rho_{2}$. Contradiction is deduced by similar reasoning as {\it Ex 1}.

{\bf Conjecture}: If two EB maps are not perfectly distinguishable in the single-copy case, then they are not in any finite-copy case.

\subsection{Holevo quantity}
Analogously, we discuss distinguishability of quantum operations measured by Holevo quantity.

{\bf Definition} Distinguishability of an ensemble of quantum operations $E=\{{\cal E}_{i}, p_{i}\}$ is defined as
\be
D(E)=\max_{\rho}\chi(\{{\cal E}_{i}\otimes I(\rho), p_{i}\}).\label{DE}
\ee

First we simplify the minimization process then show that $D(E)$ is really non-increasing under further quantum operation.

{\bf Lemma 2} Given an ensemble of quantum operations $E=\{{\cal E}_{i}, p_{i}\}$ on $d$-dimensional Hilbert space, the optimal state to achieve the distinguishability $D(E)$ is a pure state in $d\otimes d$ Hilbert space,
\be
D(E)=\max_{|\phi\rangle\in d\otimes d}\chi(\{{\cal E}_{i}\otimes I(\phi), p_{i}\}).\label{DO}
\ee
{\it Proof.} The proof is similar to that of {\it Lemma 1}. First a general mixed state $\rho^{AB}$ on $d\otimes n$-dimensional space can always be purified into a pure state $|\Phi\rangle^{ABC}$ with another auxiliary system $C$, expressed in its Schmidt decomposition splitting between $A$ and $BC$ as, \be |\Phi\rangle^{ABC}=\sum_{i=0}^{d-1}\sqrt{\lambda_{i}}|i\rangle^{A}\otimes |e_{i}\rangle^{BC}, \ee where $|e_{i}\rangle^{BC}$ are orthogonal states spanning $d$-dimensional space. From $Eq.$(\ref{trace}), we can get \be \chi(\{{\cal E}_{i}^{A}\otimes I^{B}(\rho^{AB}), p_{i}\})=\chi(\{tr_{C}{\cal E}_{i}^{A}\otimes I^{BC}(\Phi^{ABC}), p_{i}\})\le\chi(\{{\cal E}_{i}^{A}\otimes I^{BC}(\Phi^{ABC}), p_{i}\}). \ee Notice that $\Phi^{ABC}$ can be transform to the standard form $\sum_{i=0}^{d-1}\sqrt{\lambda_{i}}|i\rangle^{A}\otimes |i\rangle^{BC}$ under unitary operator $I^{A}\otimes U^{BC}$ without varying $\chi(\{{\cal E}_{i}^{A}\otimes I^{BC}(\Phi^{ABC}), p_{i}\})$. So the minimization can be considered over pure states in the $d\otimes d$-dimensional Hilbert space. For simple notation, the identity operation is omitted if no confusion appears.

{\bf Proposition 2} The distinguishability $D(E)$ is non-increasing under further quantum operations, \be D({\cal N}\circ E)\le D(E), \ee where ${\cal N}\circ E=\{{\cal N}\circ{\cal E}_{i}, p_{i}\}$ and ${\cal N}$ is a quantum operation.

{\it Proof.} The proof comes from ${\it Eq.}(\ref{operation})$ and {\it Lemma 2}. Suppose the optimal pure state to achieve $D({\cal N}\circ E)$ is $\phi^{*}$.
\bea
D({\cal N}\circ E)=\chi(\{{\cal N}({\cal E}_{i}(\phi^{*})), p_{i}\})\le\chi(\{{\cal E}_{i}(\phi^{*}), p_{i}\})\le D(E)
\eea

Additivity is a desirable property. In fact, Holevo quantity describes the asymptotic property of $\{\rho_{i}, p_{i}\}^{\otimes N}$. Just as Holevo quantity is additive on tensor product of ensembles, it is reduced to the one-copy form. How about the distinguishability defined by ${\it Eq.}$(\ref{DO})? For two ensembles of quantum operations $E_{1}=\{{\cal E}_{i}^{Q_{1}}, p_{i}\}$ and $E_{2}=\{{\cal E}_{j}^{Q_{2}}, q_{j}\}$ where $E_{1}$ and $E_{2}$ operate on $d_{1}$-dimensional system $Q_{1}$ and $d_{2}$-dimensional system $Q_{2}$ respectively, the tensor product ensemble is defined as $E_{1}\otimes E_{2}=\{{\cal E}_{i}^{Q_{1}}\otimes{\cal E}_{j}^{Q_{2}}, p_{i}q_{j}\}$. Additivity does'nt explicitly hold since for tensor product of two ensembles, the input state is possibly entangled between two ensembles to improve distinguishability of the tensor ensemble. As suggested in \cite{Childs,Acin1,D'Ariano1}, an input state entangled with an auxiliary system indeed improves distinguishability between two unitary operators. Also, entangled state can be utilized to better estimate an unknown quantum channel \cite{Acin2,D'Ariano2}. However, we show that additivity still holds. A direct corollary of {\it Proposition 2} for two ensembles of operations in sequence is as follows.

{\bf Proposition 3} For two ensembles of quantum operations $E=\{{\cal E}_{i}, p_{i}\}$ and $F=\{{\cal F}_{i}, q_{i}\}$,
\bse
\bea
D(E\circ F)\le D(E)+D(F),\\
D(F\circ E)\le D(E)+D(F),
\eea
\ese
where $E\circ F=\{{\cal E}_{i}\circ {\cal F}_{j}, p_{i}q_{j}\}$ and $F\circ E=\{{\cal F}_{i}\circ {\cal E}_{j}, q_{i}p_{j}\}$.

{\it Proof.} Suppose the optimal state to achieve $D(E\circ F)$ is $|\Phi^{*}\rangle$. \bea D(E\circ F)&=&S(\sum_{ij}p_{i}q_{j}{\cal E}_{i} \circ {\cal F}_{j}(\Phi^{*}))-\sum_{ij}p_{i}q_{j}S({\cal E}_{i} \circ {\cal F}_{j}(\Phi^{*}))
\nonumber\\
&=&S(\sum_{i}p_{i}{\cal E}_{i}(\sum_{j}q_{j}{\cal F}_{j}(\Phi^{*})))-\sum_{i}p_{i}S({\cal E}_{i}(\sum_{j}q_{j}{F}_{j}(\Phi^{*})))
\nonumber\\
&+&\sum_{i}p_{i}S({\cal E}_{i}(\sum_{j}q_{j}{\cal F}_{j}(\Phi^{*})))-\sum_{i}p_{i}\sum_{j}q_{j}S({\cal E}_{i}({\cal F}_{j}(\Phi^{*})))
\nonumber\\
&\le &D(E)+\sum_{i}p_{i}D({\cal E}_{i}\circ F)\le D(E)+D(F). \eea The second inequality is proved similarly.

{\bf Proposition 4} Distinguishability is additive on tensor product of ensembles of quantum operations
\be
D(E_{1}\otimes E_{2})=D(E_{1})+D(E_{2}). \label{DA}
\ee

{\it Proof.} From the definition of $D$ and the additivity of Holevo quantity, it is easy to show $D(E_{1}\otimes E_{2})\ge D(E_{1})+D(E_{2})$. Notice that $E_{1}\otimes E_{2}=(E_{1}\otimes I)\circ (I\otimes E_{2})$, and  $D(E_{1}\otimes E_{2})\le D(E_{1})+D(E_{2})$ holds by {\it Proposition 3}.

Now we just know that the Holevo quantity of an ensemble of quantum operations defined in {\it Eq.}(\ref{DE}) satisfies the constraint of distinguishability, therefore it is a reasonable one. But what's the physical meaning of this quantity? Can it be explained from information-theoretic point in the same way as that of quantum states? The answer is YES. It indeed represents classical information communicated by the ensemble of operations.

\subsection{Classical Information}
Before explanation, we briefly review Holevo quantity of an ensemble of quantum states since the reasoning is similar. $\chi(\{\rho_{i}, p_{i}\})$ is explained as the classical information that can be conveyed by quantum states as signals \cite{Schumacher3}. The techniques of block-coding and codeword-pruning are employed in the code-decode process \cite{Schumacher3}. More precisely, the sender encodes classical messages into long strings of states---codewords;  codewords assigned with probability $P_{s}$ (in fact they are equal) are selected from the the set of strings $\{\rho_{i_{1}}\otimes \rho_{i_{2}}\otimes\cdots\rho_{i_{n}}\}$ satisfying that the codewords are sufficiently distinguishable and the frequency of each letter respects it probability; the receiver decodes the received codeword as a whole. The maximal number of codewords is almost $2^{\chi}$. This is the information-theoretic explanation of $\chi$. We would like to explain $\chi({E})$ of an ensemble of quantum operations similarly. In this situation, quantum operations act as signal carriers instead of quantum states. The classical messages are encoded into a sequence of black boxes and each box can be inquired only once. We assert that the classical information  conveyed by an ensemble of operations is its Holevo quantity.

{\bf Proposition 5}: Let $\{{\cal E}_{i}, p_{i}\}$ be an ensemble of quantum operations acting on $d-$dimensional Hilbert space. The achievable classical information conveyed by the ensemble is
\be
I(\{{\cal E}_{i}, p_{i}\})=\max_{\phi\in d\otimes d }\chi(\{{\cal E}_{i}(\phi), p_{i}\}).
\ee

Before the proof, we emphasize that the code-decode process of quantum operations is distinct from that of quantum states though it seems alike. One different point is that a sequence of black boxes can operate in different forms of ${\cal E}_{i_{1}}\otimes{\cal E}_{i_{2}}\circ{\cal E}_{i_{3}}\cdots{\cal E}_{i_{n}}$, where $\otimes$ and $\circ$ can be varied. The other point is that the receiver has the privilege to choose a suitable input state to detect which message the sequence represents. Any entangled state is a choice so that the output state may be entangled among subsystems, i. e. optimization over entangled detector states should be considered. These two facts don't appear in the code-decode process of quantum states. However, complexity is completely solved by {\it Proposition 3, 4}.

{\it Proof.} $I \ge \chi$: The sequence of operations performs as ${\cal E}_{i_{1}}\otimes\cdots\otimes{\cal E}_{i_{n}}$ and the optimal detector state can be chosen as product state as showed by {Proposition 4}. Then the problem is reduced to that of quantum states by additivity {\it Eq.}(\ref{DA}).

$I\le \chi$: Suppose there exist a code with N codewords ${\cal E}_{i_{1}}\otimes{\cal E}_{i_{2}}\circ{\cal E}_{i_{3}}\cdots{\cal E}_{i_{n}}$ with probability $P_{s}$ such that (1) they can be almost distinguishable by $\psi$ and (2) ${\cal E}_{i}$ appears with frequency $p_{i}$. Holevo quantity is an upper bound of the classical information of $\{{\cal E}_{i_{1}}\otimes{\cal E}_{i_{2}}\circ{\cal E}_{i_{3}}\cdots{\cal E}_{i_{n}}(\psi), p_{i_{1}}p_{i_{2}}\cdots p_{i_{n}}\}$. In addition to {\it Proposition 3,4}, we have
\be
nI \le \chi(\{{\cal E}_{i},p_{i}\}\otimes\{{\cal E}_{i},p_{i}\}\circ\{{\cal E}_{i},p_{i}\}\cdots\{{\cal E}_{i},p_{i}\}(\psi))
\le n\max_{\phi\in d\otimes d }\chi(\{{\cal E}_{i}(\phi), p_{i}\}).
\ee
The proof is completed.

Now we show that $\max_{\phi\in d\otimes d }\chi(\{{\cal E}_{i}(\phi), p_{i}\})$ describes the classical information conveyed by quantum operations. The capacity of classical information conveyed by a set of states $\{\rho_{i}\}$ is defined as $C(\{\rho_{i}\})=\max_{p_{i}}\chi(\{\rho_{i},p_{i}\})$. Analogously, we can define the counterpart of a set of operations.

{\bf Definition} The capacity of classical information conveyed by a set of operations $\{{\cal E}_{i}\}$ is
\be
C(\{{\cal E}_{i}\})=\max_{p_{i}}I(\{{\cal E}_{i},p_{i}\})=\max_{p_{i}}\max_{\phi\in d\otimes d }\chi(\{{\cal E}_{i}(\phi), p_{i}\}).
\ee

{\bf Proposition 6} $C(\{\rho_{i}\}\otimes\{\sigma_{i}\})=C(\{\rho_{i}\})+C(\{\sigma_{i}\})$.

{\it Proof.}
\bea
\chi(\{\rho_{i}\otimes\sigma_{j},p_{ij}\})&=&S(\sum_{ij} p_{ij}\rho_{i}\otimes\sigma_{j})-\sum_{ij}S(\rho_{i}\otimes\sigma_{j})\nonumber\\
&\le& S(\sum_{i}p_{i|\cdot}\rho_{i})+S((\sum_{j}p_{\cdot |j}\sigma_{j}))-\sum_{i}p_{i|\cdot}S(\rho_{i})-\sum_{j}p_{\cdot |j}S(\sigma_{j}),\nonumber\\
&=&\chi(\{\rho_{i},p_{i|\cdot}\})+\chi(\{\sigma_{j},p_{\cdot |j}\}),
\eea
where $p_{i|\cdot}=\sum_{j} p_{ij}$ and $p_{\cdot |j}=\sum_{i}p_{ij}$

{\bf Proposition 7} $C(\{{\cal E}_{i}\}^{\otimes 2})=2C(\{{\cal E}_{i}\})$.

{\it Proof.} Suppose the capacity $C(\{{\cal E}_{i}\}^{\otimes 2})$ is achieved by $\{{\cal E}_{i}\otimes{\cal E}_{j}, p_{ij}\}$ and the optima input $\phi$. Then the probability of ${\cal E}_{i}$ is $(p_{i|\cdot}+p_{\cdot |i})/ 2$. According to the informational explanation of $\chi$, $\chi(\{{\cal E}_{i}\otimes{\cal E}_{j}(\phi),p_{ij}\})$ means that $2^{N\chi}$ codewords can be selected from the set of state strings $\{({\cal E}_{i_{1}}\otimes{\cal E}_{j_{1}}(\phi))\otimes\cdots\otimes({\cal E}_{i_{N}}\otimes{\cal E}_{j_{N}}(\phi))\}$ such that the frequency of ${\cal E}_{i}\otimes{\cal E}_{j}(\phi)$ is nearly $p_{ij}$ and the codewords are sufficiently distinguishable---that are measured by the average decoding error. It also amounts to that $2^{2N\cdot \chi/2}$ codewords can be selected from same set with string $2N$ such that the frequency of ${\cal E}_{i}$ occurrence is nearly $(p_{i|\cdot}+p_{\cdot |i})/ 2$ and the codewords are sufficiently distinguishable. Recall the definition of classical information of ensemble $\{{\cal E}_{i},(p_{i|\cdot}+p_{\cdot |i})/2\}$, the quantity $\chi/2$ cannot exceed $I(\{{\cal E}_{i},(p_{i|\cdot}+p_{\cdot |i})/2\})$. So we get $\chi/2\le C(\{{\cal E}_{i}\})$. The other direction is straightforward.

Comparing the classical capacity of quantum operations with one-shot capacity of classical information of a quantum channel $C_{1}({\cal E})=\max_{\{\phi_{i},p_{i}\}}\chi(\{{\cal E}(\phi_{i}), p_{i}\})$, we can see that the roles of states and channels are swapped. Also we suffer from the additivity problem.
\bse
\bea
C_{1}({\cal E}\otimes {\cal E})&=&2C_{1}({\cal E})~~~?\\
C(\{{\cal E}_{i}\}\otimes\{{\cal F}_{i}\})&=&C(\{{\cal E}_{i}\})+C(\{{\cal F}_{i}\})~~~?\\
C_{1}({\cal E}\otimes {\cal F})&=&C_{1}({\cal E})+C_{1}({\cal F})~~~?
\eea
\ese

\subsection{Unitary Ensemble}
Now we focus on the distinguishability of unitary ensemble. From the above discussion, distinguishability of unitary ensemble ${\cal U}=\{U_{i}, p_{i}\}$ acting on $d$-dimensional Hilbert space is \bea D({\cal U})=\max_{\phi\in d\otimes d}S({\cal E}(\phi)),
\nonumber\\
{\cal E}(\phi)=\sum p_{i} U_{i}|\phi\rangle\langle\phi|U_{i}^{\dagger} \label{von}. \eea Note that ${\cal E}$ can be regarded as a noisy channel though it is in some special form. Actually, a noisy channel can be expressed as this form if and only if the entanglement of assistance \cite{DiVincenzo} retains invariant after transmission through the channel \cite{Yang}. In the view of quantum state compression \cite{Schumacher2}, $D({\cal U})$ represents quantum memories required for faithfully storing the output states of the quantum channel. Universal quantum information compression \cite{Jozsa2} demonstrates that a quantum source that is only known to have von Neumann entropy less than or equal to $S$ but is otherwise completely unspecified, can be faithfully compressed to $S$ qubits by a universal quantum data compression.

First we calculate distinguishability of an ensemble $\{U_{1}, U_{2}, p_{1},p_{2}\}$ $U_{1}, U_{2}\in SU(d)$. A direct calculation shows that $S({\cal E}(\phi))=S(M)$ where $M$ is a $2 \times 2$ matrix $\left(\begin{matrix} p_{1} &\ \sqrt{p_{1}p_{2}}\langle\phi|U_{1}^{\dagger}U_{2}|\phi\rangle \nonumber\\ \sqrt{p_{1}p_{2}}\langle\phi|U_{2}^{\dagger}U_{1}|\phi\rangle &\ p_{2}\end{matrix}\right)$. The optimal state to maximize von Neumann entropy is the one to minimize $|\langle\phi|U_{1}^{\dagger}U_{2}|\phi\rangle|$ that amounts to optimally discriminating the two unitary operators. In \cite{Acin1,D'Ariano1}, minimization is achieved as follows. Taking the spectral decomposition of $U=U_{1}^{\dagger}U_{2}$ as $\{|u_{i}\rangle,u_{i}\}$, and $\rho^{A}=tr_{B}(|\phi\rangle\langle\phi|)^{AB}$, then $\langle\phi|U_{1}^{\dagger}U_{2}|\phi\rangle=tr(U\otimes I|\phi\rangle\langle\phi|)=tr\sum_{i}u_{i}|u_{i}\rangle\langle u_{i}|\rho^{A}=\sum_{i}\lambda_{i}u_{i}$, where $\lambda_{i}=\langle u_{i}|\rho^{A}|u_{i}\rangle$ is a probability distribution. In the complex plane, a polygon is formed whose vertices is the eigenvalues $u_{i}$ locating on the circle $|z|=1$. The minimum $|\sum_{i}\lambda_{i}u_{i}|$ is the shortest distance from the origin to the polygon.

It is not easy to solve the distinguishability for $SU(d)$ ensemble with more than two operators because of two obstacles. One is that the optimal states to discriminate any two operators are not the same in general, and the other is that  that even if the same optimal state minimizes all pairwise $|\langle\phi|U_{i}^{\dagger}U_{j}|\phi\rangle|$, it is unnecessarily the optimal one that maximizes von Neumann entropy---Jozsa-Schlienz paradox that occurs almost all ensembles in higher dimensions. However, we will prove the formula of distinguishability of $SU(2)$ ensemble. Fortunately, difficulties are overcome for this simple case. A maximally entangled state is optimal independently of two $SU(2)$ operators to be distinguished \cite{Acin1} and it is also the optimal state to maximize von Neumann entropy although the ensemble of output states is in $2\otimes 2$ space. As well-known, $SU(2)$ is the elementary gate for qubits that is basic system in quantum information theory. $SU(2)$ is studied with detail due to its simplicity and importance. {\it Proposition 8} is a contribution to $SU(2)$ ensemble.

{\bf Proposition 8} Distinguishability of $SU(2)$ ensemble ${\cal U}=\{U_{i},p_{i}\}$ is achieved by a maximally entangled state $|\phi^{*}\rangle=1/\sqrt{2}(|00\rangle+|11\rangle)$ and the explicit formula is,
\be
D({\cal U})=S({\cal E}(\sum_{i=1}^{n} p_{i}U_{i}|\phi\rangle\langle\phi|U_{i}^{\dagger}))=S([\sqrt{p_{i}p_{j}}trU_{i}^{\dagger}U_{j}/2]_{n\times n}).
\ee
{\it Proof.} A generic $U\in SU(2)$ matrix can be parameterized by two complex numbers $\alpha,\beta$ as \be U=\left(\begin{matrix} \alpha &\ \beta \\ -\beta^{*} &\ \alpha^{*}\end{matrix}\right) \label{uform}, \ee satisfying $|\alpha|^{2}+|\beta|^{2}=1$. Suppose the optimal state to achieve distinguishability of $\{U_{i},p_{i}\}$ is $|\psi\rangle=aV|0\rangle\otimes|0\rangle+bV|1\rangle\otimes|1\rangle$ where $a,b$ are nonnegative numbers satisfying $a^{2}+b^{2}=1$, and $V\in SU(2)$ is dependent on the ensemble (we can always fix the basis of the auxiliary subsystem in the Schmidt decomposition). Then $|\phi\rangle=a|00\rangle+b|11\rangle$ is the optimal state for $\{U_{i}V, p_{i}\}$ and $D(\{U_{i},p_{i}\})=D(\{U_{i}V, p_{i}\})$. Therefore the problem is reduced to the ensemble $\{U_{i}V, p_{i}\}$ with optimal state of the form $|\phi\rangle=a|00\rangle+b|11\rangle$. Here $V$ is certainly dependent on $\{U_{i},p_{i}\}$ and still unknown. However, we will show that $|\phi\rangle=1/\sqrt{2}(|00\rangle+|11\rangle)$ is the universal optimal state for $\{U_{i}V, p_{i}\}$ independent of $V$.

Set $U_{i}V=W_{i}\in SU(2)$ and $W_{i}$ is parameterized by $\alpha_{i},\beta_{i}$ as the form ${\it Eq.}(\ref{uform})$. The ensemble of the output states is $\{|\phi_{i}\rangle,p_{i}\}$ where \be |\phi_{i}\rangle=W_{i}|\phi\rangle=aW_{i}|0\rangle \otimes|0\rangle+bW_{i}|1\rangle\otimes|1\rangle, \ee and the average state is $\rho=\sum p_{i}|\phi_{i}\rangle\langle\phi_{i}|$. The eigenvalues of $\rho$ are the same as the non-zero eigenvalues of matrix $M$ whose entries are defined as
\be
M_{ij}=\sqrt{p_{i}p_{j}}\langle\phi_{i}|\phi_{j}\rangle.
\ee

This conclusion \cite{Jozsa1} can be easily seen by introducing $n$ orthogonal vectors $|e_{i}\rangle$ in an auxiliary Hilbert space and considering the pure state $|\Upsilon\rangle=\sum\sqrt{p_{i}}|e_{i}\rangle|\psi_{i}\rangle$. $\rho$ and $M$ are just the two reduced states obtained by partial trace of $|\Upsilon\rangle\langle\Upsilon|$ over the first and second components respectively. It follows that the Hermite matrix $M$ has the same eigenvalues as $\rho$ .

As $W_{i}^{\dagger}W_{j}\in SU(2)$, the two diagonal elements of $W_{i}^{\dagger}W_{j}$ are conjugate numbers denoted as $\alpha_{ij}$ and $\alpha_{ij}^{*}$. $M_{ij}$ is explicitly written as \be M_{ij}=\sqrt{p_{i}p_{j}}(a^{2}\langle 0|W_{i}^{\dagger}W_{j}|0\rangle+b^{2}\langle 1|W_{i}^{\dagger}W_{j}|1\rangle)=\sqrt{p_{i}p_{j}}(a^{2}\alpha_{ij}+b^{2}\alpha_{ij}^{*}). \ee We demonstrate that maximum of $S(M)$ is achieved when $a^2=b^2=1/2$ is satisfied, i. e. $|\phi^{*}\rangle$ is a maximally entangled state. This is concluded from the following reason. If the input state is maximally entangled state $|\phi^{*}\rangle$, its corresponding matrix $G$ is
\be
G_{ij}=\frac{1}{2}\sqrt{p_{i}p_{j}}(\alpha_{ij}+\alpha_{ij}^{*})=\frac{1}{2}M_{ij}+\frac{1}{2}M_{ij}^{*},
\ee
that means $G=\frac{1}{2}M+\frac{1}{2}M^{*}$.  Notice that $M^{*}$ is also a Hermite matrix with the same eigenvalues as $M$. Therefore there exists a unitary transformation $T$ satisfying $M^{*}=TMT^{\dagger}$. Now $G$ can be written as, \be G=\frac{1}{2}M+\frac{1}{2}TMT^{\dagger}. \ee It immediately follows from Uhlmann theorem that $\lambda(G)\prec \lambda(M)$, which means that the eigenvalues of $G$ is majorized by those of $M$ \cite{Nielsen2,Bhatia}. As a result of the theory of majorization, $S(G)\ge S(M)$.

Write $G_{ij}$ explicitly, \be G_{ij}=\frac{1}{2}\sqrt{p_{i}p_{j}}(\langle 0|W_{i}^{\dagger}W_{j}|0\rangle+\langle 1|W_{i}^{\dagger}W_{j}|1\rangle)=\frac{1}{2}\sqrt{p_{i}p_{j}}trW_{i}^{\dagger}W_{j} =\frac{1}{2}\sqrt{p_{i}p_{j}}tr(U_{i}V)^{\dagger}U_{j}V= \frac{1}{2}\sqrt{p_{i}p_{j}}trU_{i}^{\dagger}U_{j}. \ee It is clear that $G$ is independent of $V$. So $|\phi^{*}\rangle$ is the universal optimal state to achieve distinguishability of $SU(2)$ ensemble and the formula is obtained.

Just as Jozsa-Schlienz paradox that appears in an ensemble of quantum states \cite{Jozsa1}, we show it also occurs for an ensemble of quantum operations for the minimal dimensional case---for $SU(2)$ ensembles. It is possible to increase distinguishability of all pairwise but to decrease the global distinguishability. Notice that the phenomenon cannot occur for ensembles of quantum states in $2$-dimensional space. The paradox in case of $SU(2)$ ensemble is illustrated as follows.

{\bf Ex 3} Consider two $SU(2)$ ensembles, ${\cal U}=\{U_{1},U_{2},U_{3},1/3,1/3,1/3\}$ and ${\cal V}=\{V_{1},V_{2},V_{3},1/3,1/3,1/3\}$ where \bea U_{1}=\left(\begin{matrix} 1 &\ 0 \\ 0 &\ 1\end{matrix}\right), U_{2}=\left(\begin{matrix} \sqrt{1/2} &\ \sqrt{1/2} \\ -\sqrt{1/2} &\ \sqrt{1/2}\end{matrix}\right), U_{3}=\left(\begin{matrix} \sqrt{1/3} &\ \sqrt{2/3}e^{-i\alpha} \\ -\sqrt{2/3}e^{i\alpha} &\ \sqrt{1/3}\end{matrix}\right)
\nonumber\\
V_{1}=\left(\begin{matrix} 1 &\ 0 \\ 0 &\ 1\end{matrix}\right), V_{2}=\left(\begin{matrix} \sqrt{1/2.1} &\ \sqrt{1.1/2.1} \\ -\sqrt{1.1/2.1} &\ \sqrt{1/2.1}\end{matrix}\right), V_{3}=\left(\begin{matrix} \sqrt{1/3.1} &\ \sqrt{2.1/3.1}e^{-i\beta} \\ -\sqrt{2.1/3.1}e^{i\beta} &\ \sqrt{1/3.1}\end{matrix}\right)\nonumber
\eea
in which $cos(\alpha)=\frac{\sqrt{3}}{4}-\frac{1}{\sqrt{2}}$ and $cos(\beta)=-\frac{1}{\sqrt{2.1\times 1.1}}$.

Now it is easy to verify
\bea
|trU_{1}^{\dagger}U_{2}/2|=1/\sqrt{2} &>& |trV_{1}^{\dagger}V_{2}/2|=1/\sqrt{2.1},
\nonumber\\
|trU_{1}^{\dagger}U_{3}/2|=1/\sqrt{3} &>& |trV_{1}^{\dagger}V_{3}/2|=1/\sqrt{3.1},
\nonumber\\
|trU_{2}^{\dagger}U_{3}/2|=1/4 &>& |trV_{2}^{\dagger}V_{3}/2|=0,
\nonumber\\
D({\cal U})\approx 1.138 &>& D({\cal V})\approx 1.118,
\nonumber
\eea
which means that each pair of ${\cal U}$ is less distinguishable than the corresponding pair of ${\cal V}$, yet as a whole ${\cal U}$ is more distinguishable than ${\cal V}$.

\section{Concluding Remarks}
\label{se:remarks} In summary, we show how to define two self-consistent measures of distinguishability under a basic property of distinguishability. One is fidelity from the intuition that the more alike the less distinguishable, and the other is Holevo quantity from the information-theoretic consideration. First, distinguishability of pure states is extended to mixed ones. Then based on distinguishability of quantum states, distinguishability of quantum operations is defined as the largest distinguishability of output states. Especially, Holevo quantity of an ensemble of quantum operations is explained from the information-theoretic point. Properties of Holevo quantity of operations are discussed. The analytic formula for computing distinguishability of $SU(2)$ ensemble is proved. With the aid of the formula, we show that Jozsa-Schlienz paradox also appears in quantum operations.



\begin{thebibliography}{99}

\bibitem{Jozsa3}
R. Jozsa, J. Mod. opt. {\bf 41}, 2315 (1994).

\bibitem{Jozsa1}
R. Jozsa and J. Schlienz, Phys. Rev. A {\bf 62}, 012301 (2000).

\bibitem{Hausladen}
P. Hausladen, R. Jozsa, B. Schumacher, M. Westmoreland, and W. K. Wootters, Phys. Rev. A {\bf 54}, 1869 (1996).

\bibitem{Nielsen1}
M. A. Nielsen, C. M. Dawson, J. L. Dodd, A. Gilchrist, D. Mortimer, T. J. Osborne, M. J. Bremner, A. W. Harrow, and A. Hines, Phys. Rev. A {\bf 67}, 052301 (2003).

\bibitem{Choi}
M. D. Choi, Linear Algebra Appl. {\bf 10}, 285 (1975).

\bibitem{Schumacher1}
B. Schumacher, Phys. Rev. A {\bf 54}, 2614 (1996).

\bibitem{Childs}
A. M. Childs, J. Preskill and J. Renes, J. Mod. Opt. {\bf 47}, 155 (2000).

\bibitem{Acin1}
A. Acin, Phys. Rev. Lett. {\bf 87}, 177901 (2001)

\bibitem{D'Ariano1}
G. M. D'Ariano, P. Lo Presti, and M. G. A. Paris, Phys. Rev. Lett. {\bf 87}, 270404 (2001).

\bibitem{Kholevo}
A. S. Kholevo, Probl. Peredachi Inf. {\bf 9}, 110 (1973) [Proble. Inf. Transm. (USSR) {\bf 9}, 31 (1973)].

\bibitem{Wehrl}
A. Wehrl, Rev. Mod. Phys. {\bf 50}, 221 (1978).

\bibitem{Lindblad}
G. Lindblad, Comm. Math. Phys. {\bf 40}, 147 (1975).

\bibitem{Uhlmann}
A. Uhlmann, Comm. Math. Phys. {\bf 54}, 21 (1977).

\bibitem{Hayden}
P. Hayden, R. Jozsa, D. Petz, and A. Winter, Commun. Math. Phys. {\bf 246(2)}, 359 (2004).

\bibitem{Uhlmann2}
A. Uhlmann, Rep. Math. Phys. {\bf 9}, 273 (1976).

\bibitem{Schumacher3}
B. Schumacher and M. D. Westmoreland, Phys. Rev. A {\bf 56}, 131 (1997).

\bibitem{HSR}
Michael Horodecki, Peter W. Shor, Mary Beth Ruskai, Rev. Math. Phys. {\bf 15}, 629 (2003).

\bibitem{D'Ariano3}
G. M. D'Ariano and P. Lo Presti, Phys. Rev. Lett. {\bf 86}, 4195 (2001).

\bibitem{Acin2}
A. Acin, E. Jane, and G. Vidal, Phys. Rev. A {\bf 64}, 050302 (2001).

\bibitem{D'Ariano2}
G. M. D'Ariano and P. Lo Presti, Phys. Rev. Lett. {\bf 91}, 047902 (2003).

\bibitem{DiVincenzo}
D. P. DiVincenzo, C. A. Fuchs, H. Mabuchi, J. A. Smolin, A. Thapliyal, A. Uhlmann, {\it Entanglement of Assistance}, quant-ph/9803033.

\bibitem{Yang}
D. Yang, in preparation.

\bibitem{Schumacher2}
B. Schumacher, Phys. Rev. A {\bf 51}, 2738 (1995); R. Jozsa and B. Schumacher, J. Mod. Opt. {\bf 41}, 2343 (1994).

\bibitem{Jozsa2}
R. Jozsa, M. Horodecki, P. Horodecki, R. Horodecki, Phys. Rev. Lett. {\bf 81}, 1714 (1998).

\bibitem{Nielsen2}
M. A. Nielsen, Phys. Rev. Lett. {\bf 83}, 436 (1999).

\bibitem{Bhatia}
R. Bhatia, {\it Matrix Analysis} (Springer-verlag, New York, 1997).

\end{thebibliography}
\end{document}